\documentclass[journal]{IEEEtran}

\usepackage{cite}      

\usepackage{graphicx}
\usepackage{amsmath}
\usepackage{amsfonts}
\usepackage{url}
\usepackage{color}
\usepackage{ulem}

\newcommand{\be}{\begin{equation}}
\newcommand{\ee}{\end{equation}}
\newcommand{\bea}{\begin{eqnarray}}
\newcommand{\eea}{\end{eqnarray}}
\newcommand{\beaa}{\begin{eqnarray*}}
\newcommand{\eeaa}{\end{eqnarray*}}
\newcommand{\ben}{\begin{enumerate}}
\newcommand{\een}{\end{enumerate}}
\newcommand{\bi}{\begin{itemize}}
\newcommand{\ei}{\end{itemize}}

\newcommand{\lip}{\langle}
\newcommand{\rip}{\rangle}
\newcommand{\uu}{\underline}

\newcommand{\opex}{Opt. Express}

\newcommand{\df}{{\rm d}}

\begin{document}

\title{Assessing the effects of mode-dependent loss in space-division multiplexed systems}

\author{Anton Andrusier,~\IEEEmembership{Student Member,~IEEE}, Mark Shtaif,~\IEEEmembership{Fellow,~OSA, Senior Member,~IEEE}\\
Cristian Antonelli,~\IEEEmembership{ Member,~IEEE}, and Antonio Mecozzi,~\IEEEmembership{ Fellow,~OSA, Fellow,~IEEE}
\thanks{Manuscript received \today}
\thanks{A. Andrusier and M. Shtaif are with the Department of Physical Electronics,
Tel Aviv University, Tel Aviv 69978, Israel. A. Mecozzi and C. Antonelli are with the Department of Physical and Chemical Sciences,
University of L'Aquila, L'Aquila 67100, Italy.}}

\markboth{}{Shell
\MakeLowercase{\textit{et al.}}: }

\maketitle

\begin{abstract}
Mode-dependent loss (MDL) is known to be a major issue in space-division multiplexed (SMD) systems. Its effect on performance is complex as it affects both the data carrying signal and the accumulated amplification noise. In this paper we propose a procedure for characterizing the MDL of SDM systems by means of standard measurements that are routinely performed on SDM setups. The figure of merit that we present for quantifying MDL incorporates the effect on the transmitted signal and the noise and is directly related to the spectral efficiency reduction.
\end{abstract}


\section{Introduction}
Mode-dependent loss (MDL) is a fundamental propagation effect responsible for limiting the capacity of space-division multiplexed (SDM) systems. Although several studies of this phenomenon have been reported \cite{Winzer, HoMDL, HoMDL2, Petermann}, a workable characterization of MDL in a way that relates an easily measurable parameter with the corresponding effect on system performance is not available yet. A major issue in the case of SDM systems is the very definition of a single MDL parameter. While in the case of single-mode fibers the magnitude of MDL is fully characterized by the power-ratio between the least and the most attenuated states of polarization, in the multi-mode case a multiplicity of parameters is required to unequivocally quantify the magnitude of MDL. In this case, the obvious extension where the MDL is quantified by the power ratio between the least and the most attenuated states in the hyperpolarization space (which  consists of all spatial and polarization modes), is incomplete and cannot be unequivocally related to performance penalty. In this paper we introduce a simple and efficient approach for quantifying the effect of MDL, which is based on the definitions of a MDL vector $\vec \Gamma$ and noise degree of coherency vector $\vec \Gamma'$, which will be introduced later in this paper. We look at the reduction in spectral efficiency caused by MDL in the case where the mode-averaged signal and noise powers at the receiver are specified. We show that in the regime of small to moderate MDL, the MDL-induced reduction in spectral efficiency per mode can be expressed as
\be \frac{C_0 - C}{2N} = \frac{\Gamma^2 - \Gamma'^2}{2\ln(2)},   \label{10}\ee
where $C$ is the spectral efficiency of the actual link, and $C_0$ is the spectral efficiency of a perfect link where the received SNR equals the ratio between the mode-averaged signal and noise powers, $N$ is the number of spatial modes and $2N$ is the total number of modes including polarizations. Throughout this paper the spectral efficiency is extracted assuming an additive Gaussian noise channel with no channel state information at the transmitter \cite{Paulraj}. The quantity $\Gamma^2$ is the square length of the MDL vector $\vec \Gamma$ and it can be extracted from the channel transfer matrix $\mathbf T$ according to
\bea \Gamma^2 = 2N \mathrm{Trace}\left[\left(\frac{\mathbf{TT}^\dagger}{\mathrm{Trace} \left\{ \mathbf{TT}^\dagger \right\} } - \frac{1}{2N} \mathbf I\right)^2\right],\label{GammofT}\eea
where $\mathbf I$ is the $2N \times 2N$ identity matrix. The term $\Gamma'^2$ is the square length of the noise coherency vector $\vec \Gamma'$ and it can be obtained from the same expression (\ref{GammofT}), with $\mathbf{TT}^\dagger$ replaced by the noise coherency matrix $\mathbf{Q}$. In Sec. \ref{Measurement} we propose a simple method for extracting these quantities from experiments, without necessitating the knowledge of the full matrices $\mathbf T$ and $\mathbf Q$. Note that in the absence of MDL, where the channel transfer matrix $\mathbf T$ is unitary and the noise coherency matrix $\mathbf Q$ is proportional to the identity, $\Gamma^2 = \Gamma'^2 = 0$, and hence $C=C_0$, as expected.

It is important to stress in this context that accounting for the non-isotropy of the accumulated noise (which is captured by the deviation of $\mathbf Q$ from a constant times the identity) is critical when considering the system impact of MDL. For example, if one modeled the noise as if it were generated at the fiber input, the noise would experience the same MDL as the signal and there would be no reduction in spectral efficiency (in this case $\Gamma' = \Gamma$). {Conversely, if the noise were assumed to be produced only at the fiber end (which is equivalent to assuming that the received noise is isotropic} \cite{HoMDL, Petermann}) then $\mathbf Q $ is proportional to the identity and $\Gamma' = 0$, so that the reduction in spectral efficiency would be overestimated considerably.  The MDL-induced spectral efficiency reduction given in Eq. (\ref{10}) is general in the sense that it does not rely on any assumptions regarding the character of mode coupling, or its statistics. In the regime of strong coupling (as assumed in \cite{HoMDL}), we show in what follows that $\langle \Gamma'^2 \rangle = \frac 1 3 \langle \Gamma^2 \rangle$, and the average spectral efficiency reduction per mode simplifies to
\be \frac{C_0 - \langle C \rangle}{2N} = \frac{\langle \Gamma^2 \rangle}{3\ln(2)}, \label{40} \ee
{smaller by $2/3$ than what would follow if isotropic received noise were assumed. In this regime $\langle \Gamma^2 \rangle$ can be related to the more familiar mean MDL parameter in decibels, as we show in Sec. \ref{Strong} (see Eq. (\ref{1100})). However this relation is not very convenient as it depends on the number of supported modes.}  Finally, since MDL can be considered ergodic in frequency \cite{Antonelli}, and since the MDL correlation bandwidth is expected to be significantly narrower than the transmission bandwidth in optical fibers \cite{HoMDL2}, {the overall  MDL-induced capacity reduction per mode is obtained by multiplying Eq.(\ref{40}) by the WDM bandwidth}. In what follows, we formulate the general theory of mode-dependent loss in multi-mode fibers and obtain Eqs. (\ref{10})--(\ref{40}), which relate the MDL parameters to the information capacity of the link.

The paper is organized as follows. In Sec. \ref{Capacity} we rigorously define the vectors $\vec \Gamma$ and $\vec \Gamma'$ as well as other relevant quantities, and relate them to the channel capacity. Section \ref{Evolution} is devoted to the formulation of the equations governing the evolution of MDL and the noise degree of coherence. In Sec. \ref{Small} we apply our results to characterize the relevant regime of small MDL in conjunction with strong mode coupling. In Sec. \ref{Numerics} we validate our analysis by comparison of the theoretical results with computer simulations. In Sec. \ref{Measurement} we propose a simple method for measuring $\Gamma$ and $\Gamma'$ in SDM systems. Finally, Sec. \ref{Conclusions} is devoted to conclusions.

\section{Capacity of MDL-impaired systems \label{Capacity}}
Our analysis relies on the generalized Stokes representation of multi-mode propagation whose details have been presented in \cite{Antonelli}. The main idea is that any Hermitian matrix $\mathbf H$ of dimension $2N\times 2N$, can be conveniently expanded in a basis consisting of $4N^2-1$ matrices $\Lambda_j$ to which we refer as the generalized Gell-Mann matrices supplemented by the Identity $\mathbf I$, namely $\mathbf H = h_0\mathbf I + \sum_{j=1}^{4N^2-1}h_j\Lambda_j$, where the coefficients $h_j$ ($j=0,\dots,4N^2-1$) are real-valued. A convenient short-hand notation for the same expansion is $\mathbf H = h_0\mathbf I + \vec h\cdot \vec\Lambda$. {The properties of the matrices $\Lambda_j$ have been discussed at length in} \cite{Antonelli}. Most relevant for the analysis that follows are $\mathrm{Trace}(\Lambda_j) = 0$ and $\mathrm{Trace}(\Lambda_j\Lambda_k) = 2N\delta_{j,k}$ (here $\delta_{j,k}$ is the Kronecker delta), which imply that $\mathrm{Trace}\left[(\vec h\cdot\vec\Lambda)^2\right] = 2Nh^2$, where by $h$ we denote the modulus of the vector $\vec h$.

We consider a system in which the relation between the transmitted and received fields is given by
\bea \uu r = \mathbf T\uu x + \uu n,\label{N50}\eea
where we use underscores to denote $2N$-dimensional column vectors and bold-face letters to denote matrices (of a corresponding $2N\times 2N$ dimension). The components of $\uu r$ are the received electric fields in the various SDM modes, $\mathbf T$ is the transfer matrix of the entire channel, $\uu x$ is the vector of transmitted electric fields and $\uu n$ is the noise fields vector. The noise vector is assumed to be Gaussian and its coherency matrix is denoted by $\mathbf Q=\lip\uu n\,\uu n^\dagger\rip$. The spectral efficiency of a system defined by Eq. (\ref{N50}) is given by \cite{Paulraj}
\bea C = \log_2 \left[\det\left(\mathbf I + S \mathbf Q^{- 1/2}\mathbf T \mathbf T^\dagger \mathbf Q^{- 1/2} \right)\right],\label{N60}\eea
where $S$ is the average power transmitted in each of the components of $\uu x$. In the limit of large signal-to-noise ratio, which is appropriate in most cases of practical interest,  one can approximate $\det\left(\mathbf I + S \mathbf Q^{- 1/2}\mathbf T \mathbf T^\dagger \mathbf Q^{- 1/2} \right) \simeq \det\left(S \mathbf T \mathbf T^\dagger \mathbf Q^{-1} \right)$. Since both $\mathbf T \mathbf T^\dagger$ and $\mathbf Q$ are Hermitian, they can be expressed as
\bea \mathbf T \mathbf T^\dagger = \gamma_0\left( \mathbf I + \vec \Gamma \cdot \vec \Lambda \right),\,\, \mathbf Q = \gamma_0'\left( \mathbf I + \vec \Gamma' \cdot \vec \Lambda \right), \label{N70}\eea
where $\vec\Gamma$ is what we call the MDL vector and $\vec \Gamma'$ describes the effect of MDL on the coherency matrix of the noise. In the single-mode case \cite{Shtaif08,Anton09}, its modulus is the degree of polarization (DOP) and hence we refer to it in what follows as the noise degree of coherence (DOC) vector \cite{MecozziDOC}.
The scalars $\gamma_0$ and $\gamma_0'$ are the mode-averaged loss of the SDM system, and the mode-averaged power spectral density of the noise at the link output.
Given the matrices $\mathbf T$ and $\mathbf Q$, the components of the vectors $\vec \Gamma $ and $\vec \Gamma'$ are obtained through
\be \Gamma_n = \frac{ \mathrm{Trace} \left(  \Lambda_n \mathbf T \mathbf T^\dagger\right)}{ \mathrm{Trace} \left(  \mathbf T \mathbf T^\dagger\right) }, \,\,\, \Gamma'_n = \frac{ \mathrm{Trace} \left(  \Lambda_n \mathbf Q\right)}{ \mathrm{Trace} \left(  \mathbf Q\right) },  \label{20}\ee
whereas the scalars $\gamma_0$ and $\gamma_0'$ can be extracted from $\gamma_0 = \mathrm{Trace} \left\{ \mathbf T \mathbf T^\dagger \right\}/2N$ and $\gamma_0' = \mathrm{Trace} \left\{ \mathbf Q \right\}/2N$.

\section{The evolution of MDL and of the noise coherency matrix \label{Evolution}}
We now derive the equations governing the evolution of the vectors $\vec \Gamma$ and $\vec \Gamma'$, as well as the scalars $\gamma_0$ and $\gamma_0'$, which we subsequently apply to the regime of strong mode-coupling in order to obtain Eq. (\ref{40}).   We start by expressing the evolution equation for the transfer matrix $\mathbf T$ describing linear multi-mode propagation in a generic fiber
\bea \frac{\df {\mathbf T}}{\df z}=\frac{i}{2N}\left(\beta_0 \mathbf I+\vec\beta\cdot\vec\Lambda\right)\mathbf T  - \frac 1 2 \left[(\alpha_0-g) \mathbf I+\vec\alpha\cdot\vec\Lambda\right]\mathbf T, \label{50}\eea
where $\beta_0(z)$ and $\alpha_0(z)$ are the mode-averaged propagation constant and loss coefficient, respectively. The term $g(z)$ represents mode-independent amplification provided by in line amplifiers along the link. The real valued $(4N^2-1)$-dimensional vectors $\vec \beta(z)$ and $\vec \alpha(z)$ are, respectively, the generalized birefringence vector introduced in \cite{Antonelli} and the local mode-dependent loss vector, which we introduce here. In the single-mode case, $\vec \alpha(z)$
reduces to the local PDL vector, whose properties are discussed in \cite{ShtaifStokes}. Defining $\vec\gamma = \gamma_0\vec\Gamma$ and using Eq. (\ref{50}) we obtain
\bea \frac{\df \gamma_0}{\df z}\mathbf I + \frac{\df \vec \gamma}{\df z}\cdot\vec\Lambda &=&-(\alpha_0-g)\gamma_0\mathbf I
-\frac{1}{2}\left\{\vec\alpha\cdot\vec\Lambda,\vec \gamma\cdot\vec\Lambda\right\}\nonumber\\
&+&\frac{i}{2N}\left[\vec\beta\cdot\vec\Lambda,\vec \gamma\cdot\vec\Lambda\right]-(\alpha_0-g)\vec \gamma\cdot\vec\Lambda\nonumber\\
&-&\gamma_0\vec\alpha\cdot\vec\Lambda\label{60},\eea
where the square brackets $[\mathbf{A,B}]=\mathbf{AB-BA}$ denote the commutator of two matrices and the curly brackets $\{\mathbf{A,B}\}=\mathbf{AB+BA}$ denote their anti-commutator. To proceed we need the commutation and anti-commutation properties of the Gell-Mann matrices $\Lambda_j$. The hermiticity of these matrices implies that they have a skew-Hermitian commutator (i.e. $[\Lambda_j,\Lambda_k]^\dagger=-[\Lambda_k,\Lambda_j]$) and a Hermitian anti-commutator ($\{\Lambda_j,\Lambda_k\}^\dagger=\{\Lambda_k,\Lambda_j\}$. Hence they can each be represented as a linear combination of the matrices $\Lambda_i$,
\bea [\Lambda_j,\Lambda_k]&=&-2Ni\vec f_{jk}\cdot\vec\Lambda  \label{70}\\
\{\Lambda_j,\Lambda_k\}&=&2 \delta_{jk}\mathbf I+2N\vec q_{jk}\cdot\vec\Lambda\label{80}\eea
with the vectors $\vec f_{jk}$ and $\vec q_{jk}$ having real-valued elements, which are known in the group-theory jargon as \emph{structure constants} \cite{GroupTheory}. They are determined by the particular choice of Gell-Mann matrices and are given by
\bea f_{jk,n}&=&\frac{i}{(2N)^2}\mathrm{Trace}\big(\Lambda_n[\Lambda_j,\Lambda_k]\big) \label{90}\\ q_{jk,n}&=&\frac{1}{(2N)^2}\mathrm{Trace}\big(\Lambda_n\{\Lambda_j,\Lambda_k\}\big).\label{100}\eea
For any two vectors $\vec a$ and $\vec b$, one can define the generalized cross-product and the o-dot product as follows
\bea \vec a\times\vec b &=& \sum_{j,k}\vec f_{jk}a_jb_k \label{110}\\
\vec a\odot\vec b &=&\sum_{j,k}\vec q_{jk}a_jb_k.\label{120}\eea
In the context of multi-mode propagation, the cross-product has already been used in \cite{Antonelli} and it has the intuitive features of the familiar cross-product of three-dimensional vectors (to which it reduces for $N=1$). In particular; $\vec a\times\vec b = -\vec b\times\vec a$ and $\vec a\cdot(\vec a\times\vec b) = \vec b\cdot(\vec a\times\vec b) = 0$.  The o-dot product has been proposed in  \cite{Hu}. It is symmetric in the sense that $\vec a\odot\vec b= \vec b\odot\vec a$, and it reduces to the zero operator when the vectors $\vec a$, $\vec b$ are three-dimensional, namely for $N=1$. The cross and o-dot products of Eqs. (\ref{110}) and (\ref{120}) can be used to arrive at the following relations \cite{Antonelli,Hu}
\bea [\vec a\cdot\vec\Lambda,\vec b\cdot\vec\Lambda] &=& -2Ni(\vec a\times\vec b)\cdot\vec\Lambda\label{130}\\
\{\vec a\cdot\vec\Lambda,\vec b\cdot\vec\Lambda\} &=& 2(\vec a\cdot\vec b)\mathbf I+2N(\vec a\odot\vec b)\cdot\vec\Lambda,\label{140}\eea
which, when used in Eq. (\ref{60}), yield
\bea \frac{\df \gamma_0}{\df z} &=&-(\alpha_0-g)\gamma_0-\vec\alpha\cdot\vec \gamma\label{150}\\
\frac{\df \vec \gamma}{\df z} &=&\vec\beta\times\vec \gamma - N\vec\alpha\odot\vec \gamma -(\alpha_0-g) \vec \gamma-\gamma_0\vec\alpha. \label{160}\eea
The evolution equation for $\vec \Gamma$ is readily extracted from Eqs. (\ref{150}) and (\ref{160}) and it is given by
\bea \frac{\df \vec \Gamma}{\df z} &=& \vec\beta\times\vec\Gamma - N\vec\alpha\odot\vec\Gamma +  \vec\Gamma(\vec\alpha\cdot\vec\Gamma) - \vec\alpha. \label{170}\eea
In the case of a single-mode fiber $N=1$ and $\vec\alpha\odot\vec\Gamma=0$, so that Eq. (\ref{170}) reduces to the familiar equation for the PDL vector \cite{Huttner}. In this case, the equation is symmetric with respect to the sign of $\vec \alpha$, in the sense that if $\vec \Gamma (z)$ is a solution of Eq. (\ref{170}) for a given $\vec \alpha(z)$, then $-\vec \Gamma (z)$ solves the equation where $\vec \alpha(z)$ is replaced with $-\vec\alpha(z)$.   The term involving the o-dot product breaks this symmetry for $N\ge2$. Additionally, it can be shown that the length of the MDL vector is bounded by $\Gamma \le \sqrt{2N-1}$. Exceeding this value would contradict the positivity of the matrix $\mathbf T \mathbf T^\dagger$, as it would imply the existence of negative eigenvalues.

The evolution of the noise spectral density $\gamma_0'$ and the DOC vector $\vec\Gamma'$ is obtained by using a similar procedure. The electric field vector of the noise evolves according to
\bea \frac{\df {\uu n}}{\df z}\!\!\!\!\!&=&\!\!\!\!\!\frac{i}{2N}\left(\beta_0 \mathbf I+\vec\beta\cdot\vec\Lambda\right)\uu n- \frac 1 2 \left[(\alpha_0-g) \mathbf I+\vec\alpha\cdot\vec\Lambda\right]\uu n\nonumber\\
&&\!\!\!\!\!+ \sqrt{\hbar\omega_0 g}\,\uu s(z,t), \label{180}\eea
where $\uu s(z,t)$ accounts for additive spontaneous emission noise. We model $\uu s(z,t)$ as white noise in space. Its components satisfy $\langle s_n(z,t)s_m^*(z',t') \rangle = \delta_{m,n}\delta(z-z') R(t-t')$, where the temporal correlation function $R(t)$ is determined by the receiver frequency response, which we may assume to be matched to the signal.
Defining $\vec \gamma' = \gamma_0' \vec \Gamma'$, the equations for $\gamma_0'$ and $\vec \gamma'$ are obtained by differentiating $\langle \uu n \, \uu n^\dagger \rangle$ and by using Eq. (\ref{180}). This procedure produces the equations
\bea \frac{\df \gamma_0'}{\df z} &=&-(\alpha_0-g)\gamma_0'-\vec\alpha\cdot\vec \gamma' + \hbar\omega_0 g B\label{e13} \\
\frac{\df \vec \gamma'}{\df z} &=&\vec\beta\times\vec \gamma' - N\vec\alpha\odot\vec \gamma' -(\alpha_0-g) \vec \gamma'-\gamma_0'\vec\alpha, \label{e14}\eea
where $B=R(0)$ is the effective noise bandwidth. The equation for $\vec \Gamma'$ follows from Eqs. (\ref{e13}) and (\ref{e14}) and is given by
\bea \frac{\df \vec \Gamma'}{\df z} = \vec \beta \times \vec \Gamma' - N \vec\alpha\odot\vec\Gamma' +  \vec\Gamma'(\vec\alpha\cdot\vec\Gamma') - \vec\alpha - \vec\Gamma' \frac{\hbar\omega_0 g B}{\gamma_0'}. \label{e50}\eea
Equations (\ref{150}--\ref{170}) together with Eqs. (\ref{e13}--\ref{e50}) describe the MDL problem in a generic multi-mode link.

\section{The regime of small MDL \label{Small}}

\subsection{Capacity}
In the regime of small MDL ($\Gamma, \, \Gamma' \ll 1$), Eq. (\ref{N60}) can be approximated as
\be C = 2N \log_2\left(  \frac{S\gamma_0}{\gamma_0'} \right) - \frac{\mathrm{Trace} \left[ (\vec \Gamma \cdot \vec \Lambda)^2 - (\vec \Gamma' \cdot \vec \Lambda)^2  \right]}{2 \ln(2)},  \label{N80} \ee
where we have used the relation $\log_2 \left[\det\left(\mathbf X \right)\right] = \mathrm{Trace} \left\{ \log_2 \left(\mathbf X \right) \right\} $, as well as the fact that the determinant of a product is the product of the determinants. The first term on the right-hand side of Eq. (\ref{N80}) to the capacity of a perfect link with received signal power $S \gamma_0$ and noise    power $\gamma_0'$, which we denoted in Eq. (\ref{10}) by $C_0$. Finally, by using the property $\mathrm{Trace}\left[ (\vec h \cdot \vec \Lambda)^2 \right] = 2N h^2$ \cite{Antonelli}, Eq. (\ref{10}) is obtained. {Comparison between Eq. (\ref{10}) and the simulation results shown in Fig. 9a of} \cite{Winzer} {can be conducted only in the case where a large number of sections, each having small MDL, is considered. In this region the results shown in} \cite{Winzer} {are in agreement with Eq. (\ref{10}). Note, however, that many of the data points shown in Fig. 9a of }\cite{Winzer}{correspond to very large MDL per section, translating into extremely large link MDL (up to 80dB in power ratio between the least and the most attenuated hyper-polarizations). In this case the number of effective modes reduces considerably and a good match with the theory for small MDL presented here should not be expected.}

\subsection{Dynamic equations} In the limit of small MDL, all quantities are evaluated to first-order in the MDL vector $\vec\alpha$, and all higher order terms (which involve dot and o-dot multiplications of $\vec\alpha$ with $\vec \Gamma$ or $\vec \Gamma'$) are neglected. In this regime, a single amplified span can be approximated as an increment $\df z$ in the solution of the evolution equations, or equivalently, amplification can be viewed as distributed. The value of $g$ depends on the amplification strategy, and will typically be slightly smaller than $\alpha_0$, so as to maintain a constant output power across all modes and frequencies. When this is the case, the difference $\alpha_0-g$ can be shown to be of the order of $\Gamma^2$, which means second order in MDL, and hence we ignore it in what follows (i.e. we set $g = \alpha_0$).

Solving Eqs. (\ref{150}--\ref{170}) and (\ref{e13}--\ref{e50}) to first order in MDL, yields $\gamma_0\simeq 1$ and $\gamma_0'(z)\simeq \hbar\omega_0 gBz$. The terms $\vec\beta\times\vec\Gamma$ and $\vec\beta\times\vec\Gamma'$ appearing in Eqs. (\ref{170}) and (\ref{e50}), respectively, represent generalized rotation of the vectors $\vec\Gamma$ and $\vec\Gamma'$ about the vector $\vec\beta(z)$ in the $4N^2-1$ dimensional space \cite{Antonelli}. These terms disappear by assuming a rotating reference frame \cite{Dynamics} described by the matrix $\mathbf R(z)$ satisfying
\bea \frac{\df \mathbf R}{\df z} = \vec\beta\times\mathbf R, \eea
with the initial condition $\mathbf R(0)=\mathbf I$ and where the expression $\vec\beta\times$ should be interpreted as the matrix returning the product $\vec\beta\times \vec v$ when applied to vector $\vec v$. Denoting vectors expressed in the rotating reference frame with a tilde, namely $\vec v = \mathbf R\tilde{\vec v}$, Eqs. (\ref{170}) and (\ref{e50}) reduce to
\bea \frac{\df \tilde{\vec \Gamma}}{\df z} \simeq  - \tilde{\vec \alpha}, \hspace{0.5cm} \frac{\df \tilde{\vec \Gamma}'}{\df z} \simeq  - \tilde{\vec \alpha} -  \frac{\tilde{\vec\Gamma}'}{z}, \label{e70}\eea
whose solutions are readily shown to be
\be \tilde{\vec\Gamma} \simeq -\int_0^z \tilde{\vec\alpha}(z')\df z',\,\mbox{and } \tilde{\vec\Gamma}' \simeq -\int_0^z \frac{z'}{z}\tilde{\vec\alpha}(z')\df z'.\label{Cristian}\ee

\subsection{The regime of strong mode coupling \label{Strong}}
We now focus (as in \cite{HoMDL}) on the relevant regime of strong coupling \cite{Ho,Antonelli}, which is characterized by fast and rapidly changing $\vec\beta$, inducing similarly rapid rotations upon the vectors $\tilde{\vec\alpha}$. In this situation $\vec \alpha$ can be modeled as a delta-correlated noise, and the averages of $\Gamma^2$ and $\Gamma'^2$ can be readily calculated from the solutions of Eq. (\ref{e70}), with the result $\lip\Gamma'^2\rip = \lip\Gamma^2\rip/3$. Equation (\ref{40}) follows from the substitution of this result in Eq. (\ref{10}).

A quantity which is often used as a figure of merit in MDL analysis is the power ratio between the least and the most attenuated states of hyperpolarization, which is given by
\be \rho_{\mathrm{dB}} =  10 \log_{10} \left(  \frac{1+\lambda_{\max}}{1+\lambda_{\min}} \right) \!\simeq\! 10 \frac{\lambda_{\max} - \lambda_{\min}}{\ln(10)}, \ee
where $\lambda_{\max}$ and $\lambda_{\min}$ denote the largest and smallest eigenvalues of $\vec \Gamma \cdot \vec \Lambda$, and where the second equality holds in the regime of small MDL. Following the arguments presented in \cite{Antonelli} regarding the largest eigenvalue difference, the probability density function of $\rho_{\mathrm{dB}}$ is well approximated by a chi distribution with a shape parameter $K_N = 3+\lceil 10.39(N-1)^{1.36}  \rceil$ and a mean-square value of
\bea \langle \rho_{\mathrm{dB}}^2 \rangle = \frac{10^2}{\ln^2(10)} f(N) \langle \Gamma^2\rangle \label{1000} \eea
with
\bea  f(N) = 4 \frac{(N-1)^2 + 24.7(N-1) + 16.14 }{0.2532 (N-1)^2 + 7.401(N-1) + 16.14}. \eea
This allows us to relate the average spectral efficiency reduction per mode caused by MDL (\ref{40}) with the mean-square MDL expressed in logarithmic units,
\be \frac{C_0 - \langle C \rangle}{2N} = \frac{\ln^2(10)}{300 \ln(2) f(N)} \langle \rho_{\mathrm{dB}}^2 \rangle. \label{1100} \ee
As it is customary to refer to the mean MDL, as opposed to the root-mean square MDL that we used above, we provide the relation between $\langle \rho_{\mathrm{dB}} \rangle$ and $\langle \Gamma^2\rangle$
\bea \langle \rho_{\mathrm{dB}} \rangle = \frac{10}{\ln(10)} \frac{\mathrm{Gamma} \left(\frac{K_N+1}{2}\right) \sqrt 2}{\mathrm{Gamma} \left(\frac{K_N}{2}\right) \sqrt{K_N}}\sqrt{f(N) \langle \Gamma^2\rangle},  \label{1200} \eea
where we denote the Gamma function by ``$\mathrm{Gamma}(\cdot)$'' to avoid confusion with the modulus of the MDL vector $\Gamma$. {We stress that the analysis presented here is in accord with our claim in the introduction of this paper, challenging the suitability of $\rho_{\mathrm{dB}}$ in representing the effect of MDL in a satisfactory manner. First this is because an unequivocal relation between the capacity reduction and $\rho_{\mathrm{dB}}$ (as in Eq. (\ref{1100}) ) is only available in the regime of strong coupling, and even then it depends on the number of modes.}

\section{Numerical validation\label{Numerics}}
In our numerical procedure we concentrate only on the strong mode-coupling regime, where the MDL vector is isotropically distributed and hence the effect of birefringence (which introduces pure rotations in the generalized Stokes space) does not affect the MDL vector statistics, and need not be simulated \cite{Mecozzi}. We consider a link consisting of $ M= 100$ amplified spans, each characterized by a local MDL vector $\vec\alpha_j$ with $j=1,\dots,M$. Each vector $\vec \alpha_j$ is independently drawn at random from an isotropic Gaussian distribution with the value of $\lip\alpha^2\rip$ being a parameter. We then generate the transfer matrices $\mathbf A_j = \exp( \frac 1 2 \vec\alpha_j\cdot\vec\Lambda )$ that represents the effect of MDL in the $j$-th span such that the overall transfer matrix is $\mathbf T = \prod_{j=1}^{M}\mathbf A_j$. {The noise coherency matrix is given by $\mathbf Q = N_0 ( \mathbf I  + \sum_{k=2}^{M} \mathbf T_k \mathbf T_k^\dagger )$, where $N_0$ is a parameter representing the noise strength, and where $\mathbf T_k = \prod_{h=k}^M \mathbf A_h $ is the system transfer matrix from the beginning of the $k$-th section to the end.} Since $N_0$ does not affect the difference between the capacities with and without MDL, its value is immaterial and we arbitrarily set it to one. For each value of $\lip\alpha^2\rip$ we generated $1000$ random fiber realizations obtaining the matrices $\mathbf T$ and $\mathbf Q$ and evaluating $\vec\Gamma$, $\vec\Gamma'$ and the corresponding spectral efficiency $C$  in each realization using Eq. (\ref{N60}) for large signal-to-noise ratio.
\begin{figure}
\centering\includegraphics[width=0.9\columnwidth]{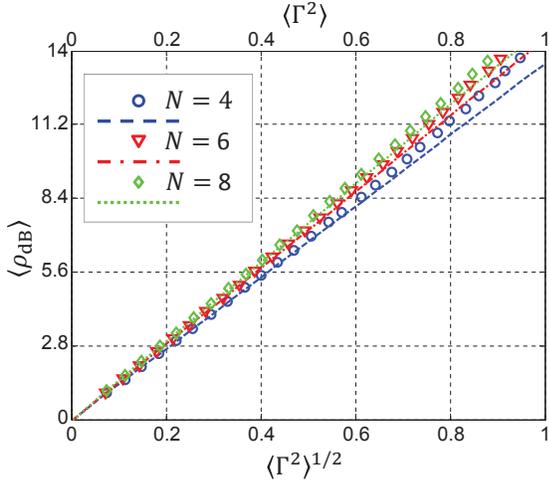}
\caption{Average MDL in decibels $\langle \rho_{\mathrm{dB}} \rangle$ versus the root-mean square value of the MDL vector length $\langle \Gamma^2 \rangle^{1/2}$. The three curves are the plot of Eq. (\ref{1200}) for $N = 4$ (dashed curve), for $N=6$ (dash-dotted  curve), and for $N=8$ (dotted curve). Circles, triangles, and diamaonds show the results of Monte Carlo simulations performed for $N=4$, $N=6$, and $N=8$, respectively. The top horizontal axis shows the values of $\langle \Gamma^2 \rangle$ which correspond to the values of $\langle \Gamma^2 \rangle^{1/2}$ in the bottom horizontal axis.}\label{Fig1}
\end{figure}

In Fig. \ref{Fig1} we show the average MDL in decibels $\langle \rho_{\mathrm{dB}} \rangle$  as a function of the root-mean square value of $\Gamma$ for different values of the number of modes $N = 2,\,4,\,6$. The top horizontal axis shows the corresponding values of the mean square value of $\Gamma$, in order to facilitate the comparison with subsequent figures. For each value of $N$ the simulations, whose results are shown by symbols in the figure, were performed for several values of $\langle \alpha^2 \rangle$, so as to explore a meaningful range of mean MDL values. The agreement with theory is self-evident. We note that even in the case of strong coupling, where the average MDL in decibels can be related to the average system penalty (Eq. (\ref{1100})), the dependence on the number of modes $N$ makes this relation less transparent. In the general case such a relation does not even exist, thereby disqualifying $\langle \rho_{\mathrm{dB}} \rangle$ from being a meaningful parameter for characterizing MDL-impaired systems.
\begin{figure}
\centering\includegraphics[width=0.9\columnwidth]{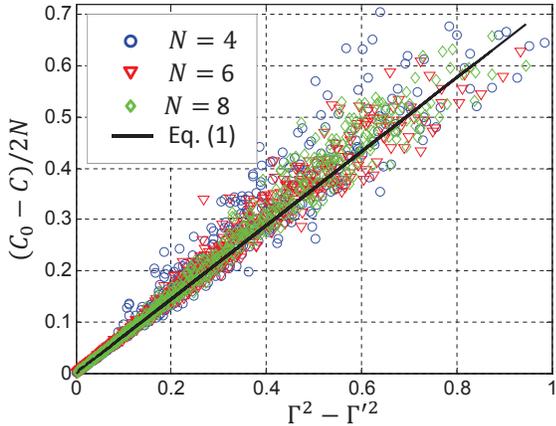}
\caption{Reduction in spectral efficiency per mode caused by MDL versus the difference $\Gamma^2-\Gamma'^2$. Symbols show individual fiber realizations randomly chosen among those used to extract the average quantities shown in Fig. \ref{Fig1}; the solid curve is the plot of Eq. (\ref{10}).}\label{Fig2}
\end{figure}

In Fig. \ref{Fig2} we test the validity of Eq. (\ref{10}) by plotting the reduction of spectral efficiency per mode as a function of the difference $\Gamma^2 - \Gamma'^2$. The symbols correspond to individual realizations of the systems that were used to extract the averaged quantities of Fig. 1. The accuracy of Eq. (\ref{10}) is evident in the entire range of simulated MDL values.
\begin{figure}
\centering\includegraphics[width=0.9\columnwidth]{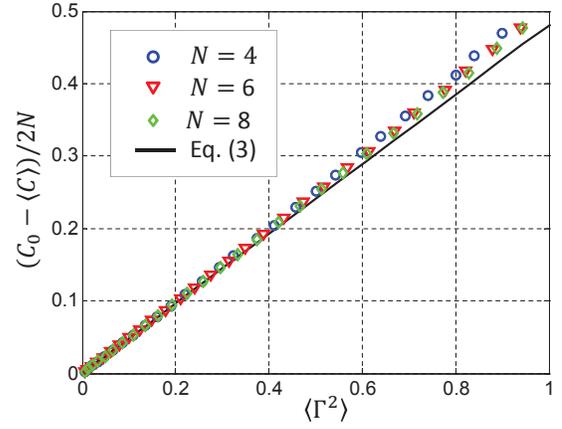}
\caption{\textit{Average} reduction in spectral efficiency per mode caused by MDL versus the mean-square MDL vector length $\langle \Gamma^2 \rangle$. Symbols show the simulation results; the solid curve is the plot of Eq. (\ref{40}). The use of the various symbols is consistent with Figs. \ref{Fig1} and \ref{Fig2}.}\label{Fig3}
\end{figure}

Finally, in Fig. \ref{Fig3} we demonstrate the accuracy of Eq. (\ref{40}), by plotting the \textit{average} reduction in spectral efficiency per mode versus $\langle \Gamma^2 \rangle/3\ln(2)$. The use of the various symbols is consistent with Figs. \ref{Fig1} and \ref{Fig2}. Once again, a very good agreement with the theoretical prediction can be observed.

\section{Extraction of the MDL parameters $\Gamma$ and $\Gamma'$ in a measurement \label{Measurement}}
\begin{figure}
\centering\includegraphics[width=0.8\columnwidth]{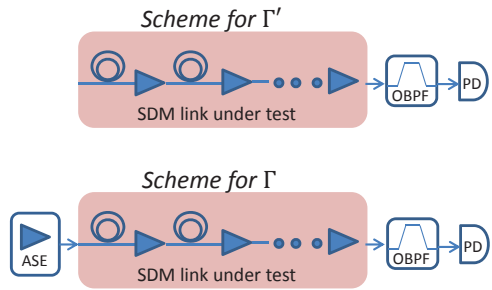}
\caption{Proposed scheme for the measurement of the two MDL parameters $\Gamma$ and $\Gamma'$. The measurement of $\Gamma$ is performed with no signal at the system input, whereas for the measurement of $\Gamma'$, a strong filtered ASE source is used as input. }\label{Fig4}
\end{figure}

While the two MDL parameters $\Gamma$ and $\Gamma'$ can be extracted from the channel transfer matrix $\mathbf T$ and from the noise coherency matrix $\mathbf Q$, it would be desirable to establish a simpler technique for their extraction that does not require the knowledge of $\mathbf T$ and $\mathbf Q$. A possible scheme of principle is illustrated in Fig. \ref{Fig4}. Since $\Gamma'$ only depends on the amplification noise statistics, its measurement is performed in the absence of an input signal. The overall noise power $P_n(t)$ (in all the modes) is measured at the link output. The value of $\Gamma'$ can be extracted from the normalized variance of the measured noise power via the relation
\be \Gamma'^2 = 2N \left( \frac{\mathrm{Var}[P_n]}{\langle P_n \rangle^2} - 1 \right), \ee
which can be shown to follow from the Gaussian statistics of the received noise. The other parameter, $\Gamma$, is measured by the exact same procedure, except that a strong thermal source (such as amplified spontaneous emission from a standard amplifier) needs to be injected into the system. The injected noise should excite all fiber modes equally and it must be much stronger than the amplification noise that is accumulated along the link. In this case
\be \Gamma^2 = 2N \left( \frac{\mathrm{Var}[P_n]}{\langle P_n \rangle^2} - 1 \right).\ee
Of course, in order to avoid undesirable averaging, the reliable extraction of the power samples requires that the received waveform be optically filtered with a bandwidth that is smaller than that of the subsequent photo-detection apparatus. The filter bandwidth should also be small relative to the modal dispersion bandwidth \cite{Antonelli}, in order to avoid frequency averaging of the measured MDL parameters.

\section{Conclusions  \label{Conclusions}}
We introduced a new approach to study MDL in spatially multiplexed systems. The key elements in this approach are the MDL vector and the noise degree-of-coherency vector, whose lengths are related in a simple form to the MDL-induced capacity reduction. These vectors can be readily extracted from measurements of the channel impulse response and noise coherency matrix. In the case of strong mode coupling, the capacity reduction comes down the mean-square length of the MDL vector divided by $3\ln(2)$.

In this paper we have studied the effect of MDL on system capacity reduction, while conditioning on the mode-averaged signal and noise powers. In practice it should be noted that these quantities are also affected nontrivially by MDL in a way that depends on the amplifiers' operation mode. The consideration of this effect is outside the scope of this work and will be left for future studies.

\section*{Acknowledgement}
\noindent This work has been carried out within an agreement funded by Alcatel-Lucent in the framework of Green Touch (\url{www.greentouch.org}). A. Andrusier and M. Shtaif acknowledge financial support from Israel Science Foundation (grant 737/12) and the Terasanta consortium. A. Mecozzi and C. Antonelli acknowledge financial support from the Italian Ministry of University and Research through ROAD-NGN project (PRIN2010-2011).


\begin{thebibliography}{99}

\bibitem{Winzer} Peter J. Winzer and Gerard J. Foschini, ``MIMO capacities and outage probabilities in spatially multiplexed optical transport systems,'' \opex {\bf19}, 16680--16696 (2011).

\bibitem{HoMDL} K-P. Ho and J.M Kahn, ``Mode-dependent loss and gain: statistics and effect on mode-division multiplexing," Opt. Express {\bf 19}, 16612--16635 (2011).

\bibitem{HoMDL2} {K-P. Ho and J.M Kahn, ``Frequency Diversity in Mode-Division Multiplexing Systems," IEEE J. Lightwave Technol. {\bf 29}, 3719--3726 (2011).}

\bibitem{Petermann} S. Warm and K. Petermann, ``Splice loss requirements in multi-mode fiber mode-division-multiplex transmission links," Opt. Express {\bf 21}, 519--532 (2013).

\bibitem{Paulraj} A. J. Paulraj, D. A. Gore, R. U. Nabar, and H. B\"{o}lcskei, ``An overview of MIMO communications -- a key to
Gigabit wireless," Proc. IEEE {\bf 92}, 198-–218 (2004).

\bibitem{Antonelli} C. Antonelli, A. Mecozzi, M. Shtaif, and P. J. Winzer,  ``Stokes-space analysis of modal dispersion in fibers with multiple mode transmission,'' Opt. Express \textbf{20}, 11718--11783 (2012).

\bibitem{Shtaif08} M. Shtaif, ``Performance degradation in coherent polarization multiplexed systems as a result of polarization dependent loss," Opt. Express {\bf 16}, 13918--13932 (2008).

\bibitem{Anton09} A. Andrusier and M. Shtaif, ``Disjoint detection in polarization multiplexed communication systems affected by polarization dependent loss," Opt. Express {\bf 17} , 8173--8174 (2009).

\bibitem{MecozziDOC} { A. Mecozzi and C. Antonelli, ``Degree of coherence in space-division multiplexed transmission,'' IEEE J. Lightwave Technol. {\bf 32}, 63--69 (2014).}

\bibitem{ShtaifStokes} M. Shtaif and O. Rosenberg, ``Polarization-Dependent Loss as a Waveform-Distorting Mechanism and Its Effect on Fiber-Optic Systems,'' IEEE J. Lightwave Technol. {\bf 23}, 923--930 (2005).

\bibitem{GroupTheory} H. Samelson, ``Notes on Lie Algebras," Springer-Verlag, (1990).

\bibitem{Hu}  Q. Hu and W. Shieh, ``Autocorrelation Function of Channel Matrix in Few-Mode Fibers with Strong Mode Coupling," to be published in Opt. Express (2013).

\bibitem{Huttner} B. Huttner, C. Geiser, and N. Gisin, ``Polarization-Induced Distortions in Optical Fiber Networks with Polarization-Mode Dispersion and Polarization-Dependent Losses,'' IEEE J. Quantum Electron. {\bf6}, 317--329 (2000).

\bibitem{Dynamics} M. Shtaif and A. Mecozzi, ``Modelling of polarization mode dispersion in optical communications systems," J. Opt. Fiber. Commun. Rep. {\bf 1}, 248–-265 (2004).

\bibitem{Ho} K-P. Ho and J.M Kahn, ``Statistics of group delays in multi-mode fibers with strong mode coupling," IEEE J. Lightwave Technol. {\bf 29}, 3119--3128 (2011).

\bibitem{Mecozzi} {A. Mecozzi and M. Shtaif, ``The statistics of polarization-dependent loss in optical communication systems,'' IEEE Photon. Technol. Lett. {\bf 14}, 313--315 (2002)}

\end{thebibliography}
\end{document}